\documentclass[10pt,letterpaper,twocolumn]{article}
\usepackage{amsfonts} 

\usepackage{ctextemp_ol2}
\usepackage[draft]{hyperref}
\usepackage{amsmath}

\begin{document}

\twocolumn[ 

\title{ Spiraling elliptic solitons in generic nonlocal nonlinear media}


\author{Guo Liang, Qian Shou and Qi Guo$^{*}$}

\address{
Laboratory of Photonic Information Technology, South China Normal
University, Guangzhou 510631,China
\\
$^*$Corresponding author: guoq@scnu.edu.cn }

\begin{abstract}We have introduced a class of spiraling elliptic solitons in generic nonlocal nonlinear media. The spiraling elliptic solitons carry the
orbital angular momentum. This class solitons are stable for any
degree of nonlocality except for the local case when the response
function of the material is Gaussian function.
\end{abstract}

\ocis{190.6135, 190.4360,060.1810.}

 ] 
\noindent Optical spatial solitons in nonlocal nonlinear media are
attracting increasing attention during recent years in both
theoretical~\cite{Snyder-science-97,Snyder-josab-97,Mitchell-josab-99,Krolikowski-pre-01,Krolikowski-pre-00,Bang-pre-02,Guo-pre-04,Ouyang-pre-06,Deng-ol-07,Deng-ol-09}
and
experimental~\cite{Peccianti-ol-02,Peccianti-apl-02,Hu-apl-06,Hu-pra-08}
aspects of research. The nonlocality plays an important role in the
nonlinear evolution of waves. It may drastically modify the
properties of solitons. The solitons in bulk Kerr media may undergo
catastrophic collapse~\cite{Moll-prl-03,Berge-pr-98}. The
nonlocality of an arbitrary shape can eliminate collapse in all
physical dimensions~\cite{Bang-pre-02}. Nonlocality can support
vortex solitons~\cite{Yakimenko-pre-06,Buccoliero-ol-08}and
multipole solitons~\cite{Buccoliero-prl-07} which are unstable in
local nonlinear media.

In theoretical aspect, ellipse-shaped solitons have been reported in
saturable nonlinear media, such as elliptic incoherent
solitons~\cite{Eugenieva-ol-00},elliptic dark
solitons~\cite{Papacharalampous-ps-04},and spiraling elliptic
solitons~\cite{Desyatnikov-prl-10}. In experimental aspect, coherent
elliptic solitons~\cite{Rotschild-prl-05} in lead glass which is
nonlocal nonlinear media and elliptic incoherent spatial
solitons~\cite{Katz-ol-04} in photorefractive sceening nonlinear
media are observed.

In this Letter, we use the variational approach to derive the
analytical spiraling elliptic solitons solution in generic nonlocal
nonlinear media. We analyze the potential function to study the
stability properties of the class of solitons.
\bigskip

The propagation of the optical beams in the nonlocal cubic nonlinear
media can be modeled by the following generic dimensionless nonlocal
nonlinear Schr$\ddot{o}$dinger
equation(NNLSE)~\cite{Mitchell-josab-99,Guo-pre-04},
\begin{equation}
i\frac{\partial \psi}{\partial
z}+\frac{1}{2}\nabla_\bot^2\psi+\Delta n\psi=0,\label{nnlse}
\end{equation}
where $\psi=\psi(x,y,z)$ is a paraxial beam, $z$ is the longitudinal
coordinate, $\nabla_\bot^2=\partial_x^2+\partial_y^2$, $x$ and $y$
are the transverse coordinates, $\Delta n=\int\int
R(x-x',y-y')|\psi(x',y',z)|{\rm d}x'{\rm d}y'$ is the normalized
nonlinear perturbation of refraction index, and $R$ is the nonlinear
response of the medium which is normalized, real and symmetric such
that$\int\int R(x,y){\rm d}x{\rm d}y=1$. We suppose the material
response to be Gaussian
function~\cite{Guo-pre-04,Krolikowski-job-04}, i.e.$R(x,y)=1/(2\pi
w_m^2)\exp[-(x^2+y^2)/2w_m^2]$, where $w_m$ is the normalized
characteristic length of the material response function.

By the variational approach~\cite{Andson-pra-83}, Eq.(\ref{nnlse})
can be interpreted as an Euler-Lagrange equation corresponding to a
vanishing variation
\begin{equation}
\delta\int\int\int
l(\psi,\psi^*,\psi_z,\psi_z^*,\psi_x,\psi_x^*,\psi_y,\psi_y^*){\rm
d}x{\rm d}y{\rm d}z=0\label{Euler-Lagrange equation},
\end{equation}
where the Lagrangian density $l$ is given
by\cite{Andson-oc-83,Guo-oc-06} {\setlength\arraycolsep{2pt}
\begin{eqnarray}
{l} &=&\frac{i}{2}(\psi^*\frac{\partial \psi}{\partial
z}-\psi\frac{\partial \psi^*}{\partial
z})-\frac{1}{2}(|\frac{\partial \psi}{\partial x}|^2+|\frac{\partial
\psi}{\partial y}|^2)\label{Lagrangian
density} \nonumber\\
&& + \frac{1}{2}|\psi|^2\int\int
R(x-\xi,y-\eta)|\psi(\xi,\eta)|^2{\rm d}\xi {\rm d}\eta.
\end{eqnarray}}
We introduce a trial function,
\begin{equation}
\psi(x,y,z)=A(z)G[X/b(z)]G[Y/c(z)]\exp(i\phi),\label{trial function}
\end{equation}
where the Gaussian envelope is $G(t)=\exp(-t^2/2)$ the phase is
$\phi=B(z)X^2+\Theta(z)XY+Q(z)Y^2+\varphi(z)$, and $X=x\cos
\beta(z)+y\sin\beta(z),Y=-x\sin\beta(z)+y\cos\beta(z)$.
Corresponding to the trial function we can obtain its power,
$P=\int\int |\psi(x,y)|^2{\rm d}x{\rm d}y=\pi A^2bc$ and orbital
angular momentum(OAM),
$M=\text{Im}\int\int\psi^*(\textbf{r}\times\nabla\psi){\rm
d}^2\textbf{r}=1/2P(b^2-c^2)\Theta$ with
$\textbf{r}=x\textbf{e}_x+y\textbf{e}_y$. Substituting the trial
function above to the variational principle Eq.(\ref{Euler-Lagrange
equation}), we obtain the reduced variational equation
\begin{equation}
\delta\int L{\rm d}z=0,\label{reduced variational equation}
\end{equation}
where $L=\int\int l_g{\rm d}x{\rm d}y$, and $l_g$ denotes the result
of inserting the Gaussian ansatz (\ref{trial function}) into the
Lagrangian density (\ref{Lagrangian density}). It also can be shown
that the Hamiltonian corresponding to Eq.(\ref{nnlse}) is of the
following form {\setlength\arraycolsep{2pt}
\begin{eqnarray}
H&=&\int\int\Bigg[\frac{1}{2}(|\frac{\partial \psi}{\partial
x}|^2+|\frac{\partial \psi}{\partial
y}|^2)-\frac{1}{2}|\psi|^2\label{Hamiltonian express} \nonumber\\
&&\int\int R(x-\xi,y-\eta)|\psi(\xi,\eta)|^2d\xi d\eta\Bigg] {\rm
d}x{\rm d}y.
\end{eqnarray}}
After some algebraic calculations, $L$ and $H$ can be analytically
determined as
{\setlength\arraycolsep{2pt}
\begin{eqnarray}
L&=&\frac{A^2 \pi }{4 b c}\Bigg[-b^2-c^2-4 b^4 B^2 c^2-4 b^2 c^4 Q^2-b^4 c^2 \Theta ^2\label{Eq7}\nonumber\\
&&-b^2 c^4 \Theta ^2+\frac{A^2 b^3 c^3 \sqrt{\left(b^2+w_m^2\right)
\left(c^2+w_m^2\right)}}{\left(b^2+w_m^2\right)
\left(c^2+w_m^2\right)}-2 b^4 c^2 B'\label{Lagrangian}\nonumber \\
&&-2 b^2 c^4 Q'+2 b^4 c^2 \Theta  \beta '-2 b^2 c^4 \Theta  \beta
'-4 b^2 c^2 \varphi '\Bigg],
\end{eqnarray}}
{\setlength\arraycolsep{2pt}
\begin{eqnarray}
H&=&\frac{A^2 \pi }{4 b c}\Bigg[b^2+c^2+4 b^4 B^2 c^2+4 b^2 c^4
Q^2+b^4 c^2 \Theta ^2\label{caculated Hamiltonian}\nonumber\\
&&+b^2 c^4 \Theta ^2-\frac{A^2 b^3 c^3 \sqrt{\left(b^2+w_m^2\right)
\left(c^2+w_m^2\right)}}{\left(b^2+w_m^2\right)
\left(c^2+w_m^2\right)} \Bigg].
\end{eqnarray}}

Following the standard procedures of the variational
approach~\cite{Andson-pra-83}, we have $b'=2bB,c'=2cQ,\beta
'=\left(b^2+c^2\right) \Theta
\left/\left(b^2-c^2\right)\right.,P'=0,H'=0$ and $M'=0$. Primes
indicate derivatives with respect to the evolution variable $z$. So
we can rewrite the Hamiltonian of the system as follows,
\begin{equation}
H=\frac{P}{4}(b'^2+c'^2+\Pi)\label{Hamiltonian},
\end{equation}
{\setlength\arraycolsep{2pt}
\begin{eqnarray}
\Pi&=&\frac{1}{b^2}+\frac{1}{c^2}+\frac{4 b^2 \sigma
^2}{\left(b^2-c^2\right)^2}+\frac{4 c^2 \sigma
^2}{\left(b^2-c^2\right)^2}\label{potential function}\nonumber\\
&&-\frac{P}{\pi  \sqrt{\left(b^2+w_m^2\right)
\left(c^2+w_m^2\right)}},
\end{eqnarray}}
with $\sigma\equiv M/P=1/2(b^2-c^2)\Theta$.

\begin{figure}[htb]
\centerline{\includegraphics[width=7.5cm]{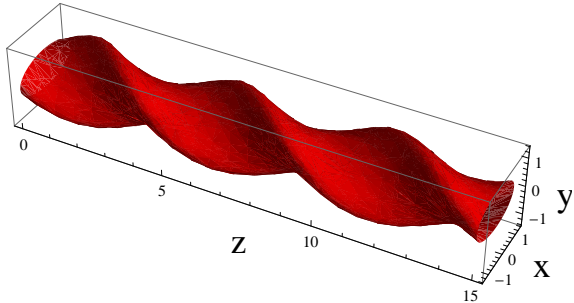}}
\caption{(color online) Propagation dynamics of the spiraling
elliptic soliton in nonlocal nonlinear media. The isointensity plot
is at the level $I_m/2$ of the elliptic soliton with $I_m=20256.03$
where $I_m$ is $\text{max}|\psi|^2$.The normalized characteristic
length of the material response function $w_m$ is 15, and the degree
of nonlocality $\alpha$ is 7.5.}\label{propagation}
\end{figure}

\begin{figure}[htb]
\centerline{\includegraphics[width=7.5cm]{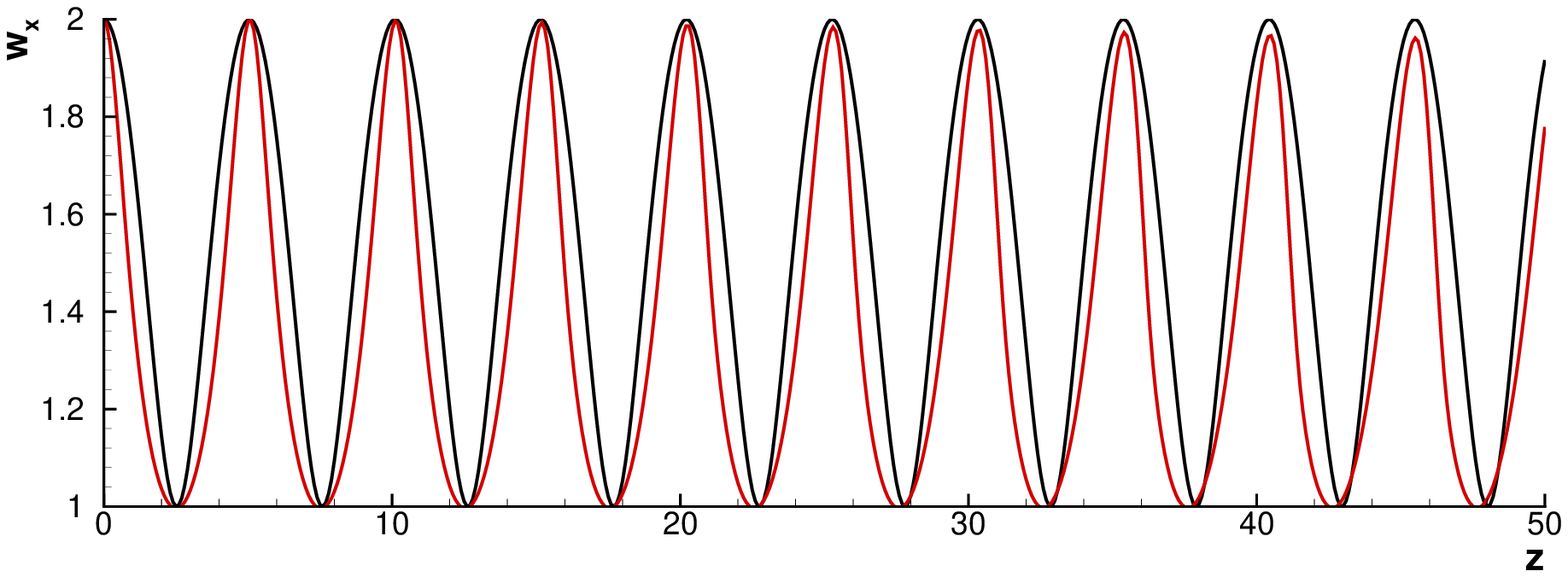}} \caption{(color
online) Evolution of the beam width of the spiraling elliptic
soliton in the direction of x axis. Numberically obtained half width
$w_x$ (black line) is compared to the variational result (red line).
The parameters $w_m$ and $\alpha$ are 15 and 7.5
respectively.}\label{wx15}
\end{figure}

\begin{figure}[htb]
\centerline{\includegraphics[width=7.5cm]{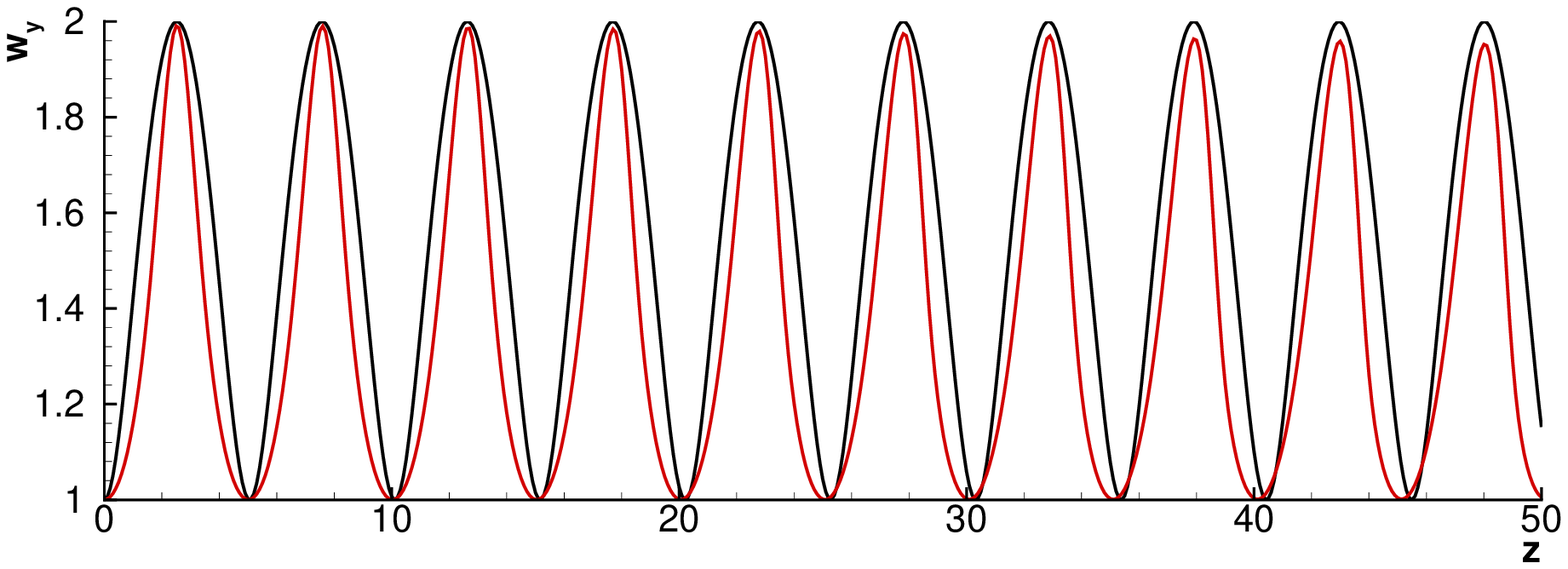}} \caption{(color
online)Same as Fig.\ref{wx15} but with plot corresponding to the
beam width of the spiraling elliptic soliton in the direction of y
axis.}\label{wy15}
\end{figure}

Solitons can be found as the extrema of the potential $\Pi(b,c)$.
Letting $\partial\Pi/\partial b=0$ and $\partial\Pi/\partial c=0$,
we can obtain the critical power and the OAM.
{\setlength\arraycolsep{2pt}
\begin{eqnarray}
P_c&=&\frac{2 \left(b^2+c^2\right)^3 \pi
\left[\left(b^2+w_m^2\right)
\left(c^2+w_m^2\right)\right]{}^{3/2}}{b^4 c^4 \left[b^4+6 b^2
c^2+c^4+4 \left(b^2+c^2\right) w_m^2\right]}\label{critical power},\\
\sigma_c^2&=&\frac{\left(b^2-c^2\right)^4 \left[b^2
c^2+\left(b^2+c^2\right) w_m^2\right]}{4 b^4 c^4 \left[b^4+6 b^2
c^2+c^4+4 \left(b^2+c^2\right) w_m^2\right]}\label{critical OAM}.
\end{eqnarray}}
We can also obtain the rotation velocity $\omega \equiv \beta '=2
\left(b^2+c^2\right) \sigma \left/\left(b^2-c^2\right)^2\right.$.
When the input power and OAM are chosen arbitrarily the spiraling
elliptic solitons can be found the semi-axis of which are determined
by Eq.{\ref{critical power}} and Eq.{\ref{critical OAM}}. One
example is shown in Fig.\ref{propagation} with $P_c=127272.4$ and
$\sigma_c=0.560949$ (other parameters are
$b=2.0,c=1.0,\Theta=0.373966,w_m=15.0$ and $\omega=0.623277$). The
isosurface of intensity of the spiraling soliton is obtained from
our variational solution. Comparing two half widths obtained from
variational solution, $w_x=\sqrt{b^2 \text{cos}^2\omega  z+c^2
\text{sin}^2\omega  z}$ and $w_y=\sqrt{c^2 \text{cos}^2\omega  z+b^2
\text{sin}^2\omega  z}$, with the numerical results we find an
excellent agreement as is shown in Fig.\ref{wx15} and Fig.\ref{wy15}
We introduce a nonlocal parameter $\alpha=\text{max}(w_m/b,w_m/c)$
to define the degree of nonlocality for the beam in nonlocal
nonlinear media. The larger is the nonlocal parameter, the stronger
is the degree of nonlocality. In Fig.\ref{propagation},
Fig.\ref{wx15} and Fig.\ref{wy15} $w_m$ is $15.0$, and the degree of
nonlocality $\alpha$ is $7.5$.

\begin{figure}[htb]
\centerline{\includegraphics[width=7.5cm]{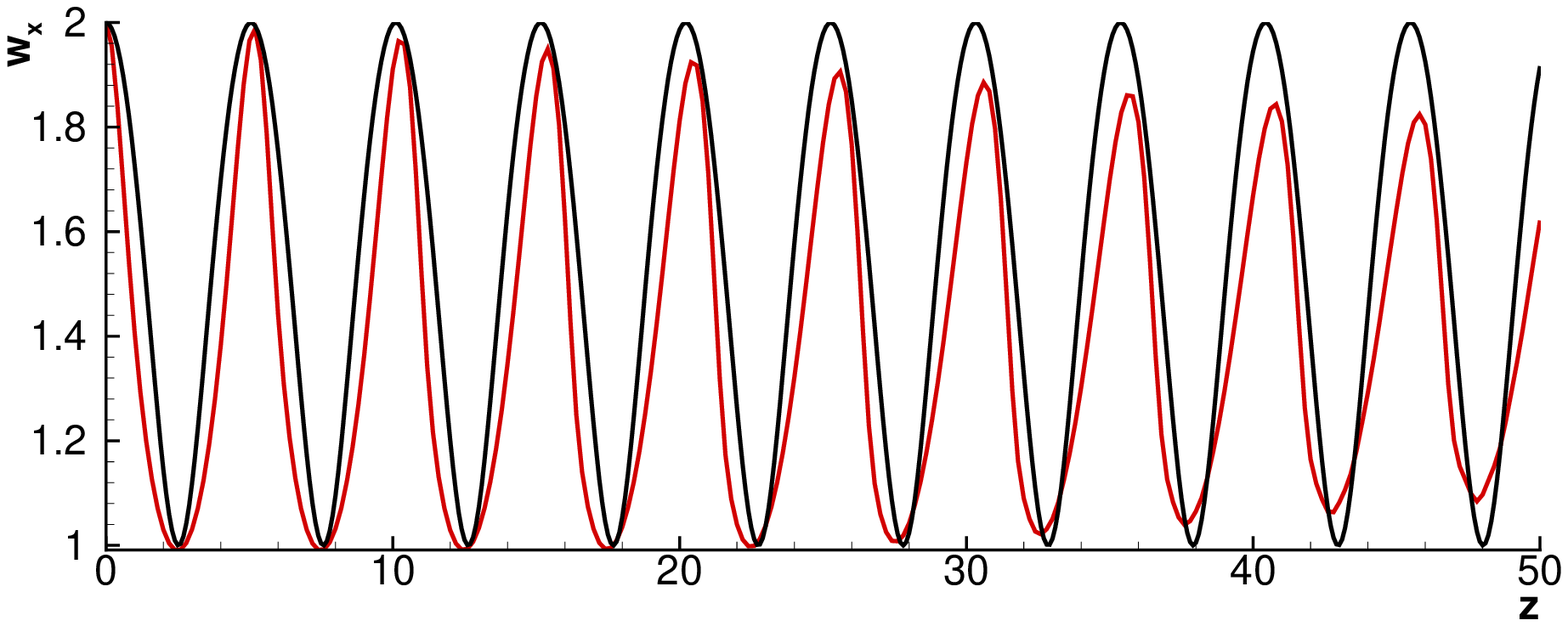}} \caption{(color
online) Evolution of the beam width of the spiraling elliptic
soliton in the direction of x axis. Numberically obtained half width
$w_x$ (black line) is compared to the variational result (red line).
The parameters $w_m$ and $\alpha$ are 8.0 and 4.0
respectively.}\label{wx8}
\end{figure}

\begin{figure}[htb]
\centerline{\includegraphics[width=7.5cm]{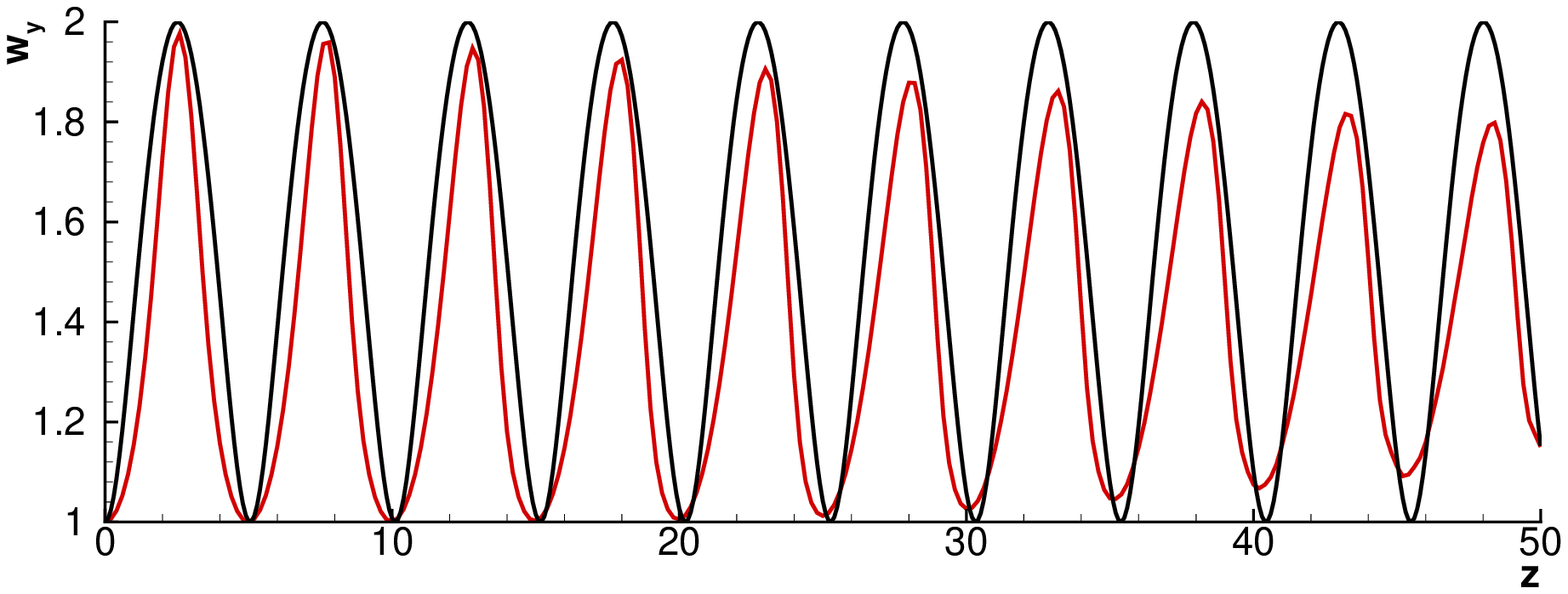}} \caption{(color
online) Same as Fig.\ref{wx8} but with plot corresponding to the
beam width of the spiraling elliptic soliton in the direction of y
axis.}\label{wy8}
\end{figure}

An important aspect of any family of soliton solutions is their
stability properties. We can study the stability characteristics of
our analytical soliton solution by means of the analysis of the
potential function $\Pi(b,c)$. So we search the second derivative of
the potential $\Pi(b,c)$ with respect to $b$ and $c$, then
substituting Eq.{\ref{critical power}} and Eq.{\ref{critical OAM}}
into it we get
 {\setlength\arraycolsep{2pt}
\begin{eqnarray}
\frac{\partial ^2\Pi }{\partial b^2}&=&\frac{2
\left(b^2+c^2\right)}{b^4c^4\left(b^2+w_m^2\right)\left[b^4+6b^2
c^2+c^4+4\left(b^2+c^2\right) w_m^2\right]}\label{potential 2nd derivative to b}\nonumber\\
&&\Bigg[b^2c^2\left(b^4+14b^2
c^2+c^4\right)+(b^6+18b^4c^2+33b^2c^4\nonumber\\
&&+4c^6)w_m^2+\left(b^4+5b^2c^2+16c^4\right)w_m^4\Bigg],\\
\frac{\partial ^2\Pi}{\partial b\partial c}&=&-\frac{2
\left(b^2+c^2\right) }{b^3 c^3 \left[b^4+6 b^2 c^2+c^4+4
\left(b^2+c^2\right) w_m^2\right]}\label{potential 2nd derivative to bc}\nonumber\\
&&\Bigg[b^4+14 b^2 c^2+c^4+12 \left(b^2+c^2\right) w_m^2\Bigg],
\end{eqnarray}
} {\setlength\arraycolsep{2pt}
\begin{eqnarray}
\frac{\partial^2\Pi}{\partial c^2}&=&\frac{2
\left(b^2+c^2\right)}{b^4 c^4 \left(c^2+w_m^2\right) \left[b^4+6 b^2
c^2+c^4+4 \left(b^2+c^2\right) w_m^2\right]}\label{potential 2nd derivative to c}\nonumber\\
&&\Bigg[b^2 c^2 \left(b^4+14 b^2 c^2+c^4\right)+\left(4 b^6+33 b^4
c^2+18 b^2 c^4\right.\nonumber\\
&&\left.\left.+c^6\right) w_m^2+\left(4 b^4+5 b^2 c^2+4 c^4\right)
w_m^4\right.\Bigg].
\end{eqnarray}
} From Eq.{\ref{potential 2nd derivative to b}}, Eq.{\ref{potential
2nd derivative to bc}} and Eq.{\ref{potential 2nd derivative to c}}
we can easily get $\partial^2\Pi/\partial b^2>0$,
$\partial^2\Pi/\partial c^2>0$ and
$\triangle\equiv(\partial^2\Pi/\partial b^2)(\partial^2\Pi/\partial
c^2)-(\partial^2\Pi/\partial b\partial c)^2>0$ when $w_m\neq0$ (the
degree of nonlocality $\alpha$ is not zero). So $b$ and $c$ of the
spiraling elliptic solitons what we have got analytically by the
variational approach are corresponding to the minima of the
potential $\Pi(b,c)$. So the soliton solutions are stable for any
degree of nonlocality except for the local case. But we should
mention that when the degree of nonlocality decreases the
low-intensity oscillating tails which indicate the appearance of
dispersive waves radiated by the soliton will
occur~\cite{Desyatnikov-prl-10,Yang-pre-02}. The radiative tails
take a portion of radiated OAM from the soliton, and the reduction
of OAM in the main soliton leads to the slow reduction of
ellipticity of the transverse rotating
profile~\cite{Desyatnikov-prl-10}. This can be verified in
Fig.\ref{wx8} and Fig.\ref{wy8}. In Fig.\ref{wx8} and Fig.\ref{wy8}
$w_m$ is $8.0$, and the degree of nonlocality $\alpha$ is $4.0$.

In conclusion, we have obtained spiraling elliptic solitons in
generic nonlocal nonlinear media by use of the variational approach.
We show that this class of solitons are stable for any degree of
nonlocality except for the local case. Because of the appearance of
dispersive waves radiated by the soliton the ellipticity of the
spiraling elliptic solitons will reduce when the degree of
nonlocality becomes lower. Our theoretical results have been
confirmed by direct numerical simulations of the NNLSE.

\pagebreak

This research was supported by the National Natural Science
 Foundation of China (Grant Nos. 11074080 and 10904041), the Specialized Research Fund for the Doctoral Program
 of Higher Education (Grant No. 20094407110008), and the Natural Science Foundation
of Guangdong Province of China (Grant No. 10151063101000017).

\end{document}